\documentstyle[manuscript,aps]{revtex}
\newcommand{\beq}{\begin{equation}}
\newcommand{\eeq}{\end{equation}}
\newcommand{\beqa}{\begin{eqnarray}}
\newcommand{\eeqa}{\end{eqnarray}}
\newcommand{\beqar}{\begin{eqnarray*}}
\newcommand{\eeqar}{\end{eqnarray*}}

\def \A {{\bf A}}
\def \U {{\bf U_\Delta}}
\def \M {{\bf M}}
\def \didnt {({\bf 1}-\M)}
\def \m {{\bf m}}

\def \J {{\bf J}}
\def \H {{\bf H}}

\def \la {\langle}
\def \ra {\rangle}

\def \up {\uparrow_z}
\def \down {\downarrow_z}
\def \Q {{\bf Q}}
\def \P {{\bf P}}

\begin{document}



\title{ \bf\Large 
When does a Measurement or Event Occur?
}

\author{ { 
J. Oppenheim$^{(a)}$,\footnote{\it jono@physics.ubc.ca}
B. Reznik$^{(b)}$,\footnote{\it reznik@t6-serv.lanl.gov}
and W. G. Unruh$^{(a)}$}\footnote{\it unruh@physics.ubc.ca \\
}  
{\ } \\
(a) {\it \small   Department of Physics,  University of British
Columbia,
6224 Agricultural Rd. Vancouver, B.C., Canada
V6T1Z1}\\
(b) {\it \small Theoretical Division, T-6, MS B288, 
Los Alamos National Laboratory, Los Alamos, NM, 87545}
}
\maketitle

\begin{abstract}
Within quantum mechanics it is possible to assign a probability to the
chance that a measurement has been made at a specific time t.
However, the interpretation of such a probability is far from clear.
We argue that a recent measuring scheme of Rovelli's (quant-ph/9802020) 
yields probabilities which do not correspond to the conventional
probabilities usually assigned in quantum mechanics.
The same arguments also apply to attempts to use the probability current
to measure the time at which a particle arrives at a given location.
\end{abstract}
\newpage
%
%
%
%
The observable $\A$ of a quantum system $S$ can be measured by
coupling a macroscopic apparatus $O$ to it, via an interaction such as
\beq
H=g(t)\P\A
\eeq
where $\P$ is the conjugate momentum to the pointer $\Q$ of the
measuring device, and $g(t)$ is a function which is zero everywhere,
except during a small interval of time.  After the measurement is
complete, the measuring apparatus will be correlated with the state
of the system.  If initially, $S$ is in a superposition of eigenstates
$|\phi_i\rangle$ of the observable $\A$, so that $|\psi_S\rangle=
\sum_i c_i |\phi_i\rangle$,
then we expect the initial state of the combined $S-O$ system to evolve
into a correlated state.
\beq
\sum_i c_i |\phi_i\rangle \otimes |O\rangle \rightarrow
\sum_i c_i |\phi_i\rangle \otimes |O_i\rangle
\eeq
where $|O\ra$ is the original state of the device and the $|O_i\rangle $ are orthogonal states of the measuring apparatus which are correlated with the 
system.
If the coupling is small, then
the duration of the measurement might need to be long in order to
distinguish between the various eigenvalues of $\A$.  At any time during
the measurement, it is possible to calculate the density matrix of the
combined S-O system .  One can imagine that
a second apparatus $O'$ measures the state of the first apparatus $O$
to determine whether a measurement has occurred.  This has been
studied for the case when the measurement is gradual \cite{pereswooters}
\cite{ar}. Rovelli
\cite{rovelli} has recently
proposed that the apparatus $O'$ might measure the operator
\beq
\M=\sum_i | \phi_i\rangle \otimes |O_i\rangle
\langle O_i | \otimes \langle \phi_i |. \label{eq:m}
\eeq
This is a projector onto the space of states where a correlation
exists between the measuring apparatus and the quantum system.
In the measurement scheme proposed by Rovelli, the probability
that a measurement has been made at time $t$ is given in the Heisenberg
representation by
\beq
P_M(t)=\langle \psi_{SO} |\M (t)| \psi_{SO} \rangle
\eeq
where $|\psi_{SO}\rangle$ is the state of the combined S-O system.
The operator which gives the probability that a measurement was made
between times $t$ and $t+dt$
is 
\beq
\m(t)=\frac{d\M(t)}{dt}.
\eeq

It is interesting to compare this method for measuring the time of a 
measurement to the use of the probability current to measure the 
time-of-arrival \cite{jad}.  One imagines that a particle is localized 
in the region $x < 0$ and travelling towards the origin.
The projector ${\bf \Pi}_+=\int_0^\infty dx |x\ra\la x|$ is an
operator which is equal to one when
$x > 0$ and zero otherwise.  The probability of detecting the
particle in the positive x-axis is given by
\beq
P_+(t)=\langle \psi |{\bf \Pi}_+(t)| \psi \rangle .
\eeq
In the Schr\"{o}dinger representation, this expression is just $P_+(t)=\int_0^\infty  |\psi(x,t)|^2dx$.
It is then claimed that the current $\J_+$, given by
\beq
\frac{\partial \J_+}{\partial x} =\frac{d{\bf \Pi}_+(t)}{dt}
\eeq
will give the probability that the particle arrives between $t$ and
$t+dt$.

The problem lies with interpreting these probabilities as
{\it probabilities in time}.  In conventional quantum mechanics, for each
observable,
we can assign an operator $\A$.  At each time $t$ there exists a Hilbert
space and 
inner product which enable one to calculate the probability 
$P_a(t)$ that an observation yields the result $a$.  The probability of 
finding the result $a$ at time $t$ is independent of
the
probability of finding the result $a'$ at the same time.  
ie. if ${\bf \Pi}_a(t)$ is the projection operator
onto the state with eigenvalue $a$, then 
\beq
[{\bf \Pi}_a(t),{\bf \Pi}_{a'}(t)]=0.
\eeq 
If $a\neq a'$ then the projection operators project onto orthogonal
states.

If we interpret the probabilities
$P_a(t)=\langle\psi|{\bf\Pi}_a(t)|\psi\rangle$ 
as probabilities in time, then our 
conventional notions of what these probabilities mean, break down.
In general, 
\beq
[{\bf \Pi}_a(t),{\bf \Pi}_a(t')] \neq 0.
\eeq
Measurements made
at earlier times influence measurements made at later times.  One doesn't
expect any of the 
operators $\M(t)$ and $\m(t)$ (or for the time-of-arrival case, 
${\bf \Pi}_+(t)$ and $\J_+(t)$) to commute with
themselves at different times
\footnote{A similar problem is encountered when on attempts to
measure the dwell time for a particle tunnelling through a potential
barrier\cite{oppenheim}}.  
Furthermore, one no longer has the same notion of orthogonality as with
conventional probabilities.  At a given time $t$ a system can only take on 
one value $a$, but for a given value $a$, a system may attain this value
at many different times $t$.  Furthermore, the probabilities $P_a(t)$,
while
they are normalized at a given time
\beq
\sum_a P_a(t) = 1
\eeq
are not normalized as probabilities in time.
\beq
\int_{-\infty}^{\infty} dt P_a(t) \neq 1
\eeq
The norm can even be zero or infinite.  One can try to normalize the
probabilities,
but the normalization is different for each state $\psi$, and one needs to
know the 
state $\psi$ at all times before the normalization can be done.

Another problem is that probabilities derived from operators such as
$\m(t)$  will be negative
when $\M(t)$ is not a monotonically increasing function of time.  
In addition, one finds that an operator such as $\J_+(t)$
can attain negative values, even for particles which only contain
modes of positive frequency \cite{backflow}. 
The possibility of negative values for these quantities makes it
impossible
to interpret them as probabilities.  
Furthermore, it can be shown \cite{aharonov} that formally, an operator
which measures the time of the occurrence of an event
cannot exist.  
\vskip .5cm
 
\noindent 
Instead of considering operators, a more physical meaningful method of 
measuring the occurrence of an event is to consider continuous
measurement processes.  For example, the operator ${\bf \Pi}_a(t)$
can be measured continuously or at small time intervals.  The probability
of finding that the system enters the state $\phi_a$ at time $t_a$
is given by the probability that it isn't in the state $\phi_a$ before
$t_a$,
times the probability that it is in the state $\phi_a$ at $t_a$.

To see how such a scheme might work, let us see how one would measure
the time of an occurrence of a measurement.  A measurement of the operator
$\M(t)$ will tell us whether a measurement has occurred. 
We can then measure $\M(t)$ at times $t_k=k \Delta$ for integral $k$
in order to determine when the measurement occurred.
We will now work in the Schr\"{o}dinger representation, simply because
it is the most natural arena to talk about successive measurements on a 
system.

At time $t_1$, the probability that a measurement has occurred is given
by
\beq
P(\up, t_1) = \la \psi_0 (0)| \U^\dagger  \M \U |\psi_0(0) \ra
\eeq
and the probability that it hasn't is
\beq
P(\down, t_1) = \la \psi_0 (0) |\U^\dagger \didnt \U |\psi_0(0) \ra
\eeq
where $\up$ and $\down$ correspond to detection or null, $\psi_0(0)$
is the initial state of the system and $\U$ is the evolution operator $e^{-i\H\Delta}$.
If the result is null, we collapse the wave function and evolve
it to the next instant. The normalized state before the second
measurement is:
\beq
|\psi_2(t_2) \ra = {\U \didnt \U |\psi_0\ra \over
          \la \psi_0|\U^\dagger \didnt \U |\psi_0 \ra^{1/2} }
\eeq
The probability that a measurement has occurred at $t_2$ is given by
the probability that a measurement didn't occur at $t_1$ times the
probability
that $\psi_2$ is in one of the states $|\phi_i\ra \otimes |O_i\ra$
\beq
P(\up, t_2) = {\la\psi_0|  \U^\dagger \didnt \U^\dagger \M
     \U \didnt \U |\psi_0\ra \over
          \la  \psi_0|\U^\dagger \didnt \U|\psi_0 \ra }
\times
 \la \psi_0|\U^\dagger\didnt\U |\psi_0 \ra
\eeq
The probability that a measurement didn't occur is given by
\beq
P(\down,t_2)=\la\psi_0|\U^\dagger \didnt \U^\dagger \didnt
     \U \didnt \U |\psi_0\ra
\eeq
By repeating this process,
we find that at time $t_k$ the probability that a measurement has occurred
is given
by
\beq
P(\up, t_k) = \la \psi_0 | A_k | \psi_0\ra
\eeq
where
\beq
A_k = \U^\dagger\didnt \U^\dagger \didnt ...\U^\dagger \M
\U ... \didnt \U \didnt \U
\eeq
and the probability that a measurement hasn't occurred is 
\beq
P(\down, t_k) = \la \psi_0 | B_k | \psi_0\ra
\eeq 
with
\beq
B_k=\U^\dagger \didnt \U^\dagger \didnt ...\U^\dagger \didnt
\U... \didnt \U \didnt \U \label{eq:nomeas}
\eeq
By acting the unitary operators on the projection
operators we can write the $A_k$ or $B_k$ in the "Heisenberg
representation."
For example 
\beqa
A_k=\didnt(t_1)...\didnt(t_{k-1})  \didnt(t_k)
\didnt(t_{k-1})...\didnt(t_1)
\eeqa
However, while the operators $\M (t)$ can be found by unitary
time-evolution of $\M(0)$, the operators $A_k$ and $B_k$ are not
related
by a unitary transformation to $A_0$ and $B_0$.  Nor are the $A_k$ and
$B_k$
projection operators. 
By construction, the sum of the probabilities, 
$\sum_{k=1}^\infty P(\up, t_k) =1$,  
however, these probabilities are not universal.
In this case, they apply only to the particular measurement scenario
under discussion.  As we will now show, the probability distribution
is sensitive to the frequency at which $\M$ is measured, a phenomenon which
is related to the Zeno paradox \cite{zeno}.  

As an example, consider a spin 1/2 particle which is in a state given by
\beq
|\psi_S\ra = a|\up\ra + b|\down\ra
\eeq
and a simple measuring device which is also a spin 1/2 particle initially
in the state $|O\ra=|\up'\ra$, which evolves according to the 
Hamiltonian
\beq
\H=g(t) \sigma_x' (1-\sigma_z)
\eeq
where $g(t)={\pi\over T}$ when $0<t<T$ and the primed  Pauli matrix acts on the measuring device, while
the unprimed Pauli matrix acts on the system.
After a time $T$, the spin of the measuring device will be correlated
with the system.  Since this measurement is rather crude, (the
initial state of the device is the same as one of the measurement states),
the operator $\M$ at $t=0$ is not zero. 
Let us simplify the problem further, by assuming that $a=0$ and $b=1$.
In this case, the only relevant matrix element of $\didnt$ is 
$|\up \ra \otimes |\down' \ra \la\down' | \otimes \la\up | = |\psi_o\ra\la\psi_o|$. 
We then find
the probability that the measuring apparatus has not responded at time 
$t_k$ is
\beqa
P(\down,t_k) & = & |\la \psi_o |U_\Delta |\psi_o\ra |^{2k} \nonumber\\
&\simeq&  |\la \psi_o | 1-i\Delta \H - \Delta^2 \H^2 |\psi_o\ra |^{2k} \nonumber\\
&\simeq& 1-\Delta^{2k}(\la \H^2 \ra-\la \H \ra ^2)^{k} 
\eeqa
If we fix a value of $\tau=t_k$ and then make $\Delta $ go to zero, 
we find
\beqa
P(\down,\tau) & \simeq & e^{-(\Delta dE)^{2\tau/\Delta}} \nonumber\\
& \simeq & 1,
\eeqa 
which implies that the measuring apparatus becomes frozen and never
records a measurement.
In order not to freeze the apparatus, we need
$\Delta >  1/dE $.  There is always an inherent inaccuracy when measuring
the time that the event (of the measurement) occurred.
This inaccuracy is similar to the one found when trying to measure
the time-of-arrival  or the traversal-time \cite{aharonov} \cite{oppenheim}
\cite{peres} \cite{allcock}. 
Note that this inaccuracy is not related to the so-called "Heisenberg
energy-time uncertainty relationship" as it applies to every single
measurement and not to the width of measurements carried out on an ensemble.

\vskip .5cm
\noindent We have seen that operators which
classically might give the time of an event cannot be given a
physical interpretation.  Several authors \cite{grot}\cite{muga}
have maintained that the  
problems with defining an operator for the time of an event are 
technical, and can be circumvented by slightly 
modifying these operators.  However, we have argued that probabilities
in time are fundamentally different from traditional probabilities
in Quantum Mechanics, and that there is a limitation on these measurements.



\end{document}